\documentclass[aps,pra,preprint,superscriptaddress,floatfix,longbibliography]{revtex4-1}
\usepackage{dcolumn}
\usepackage{graphicx}
\usepackage{amsmath,amssymb}
\usepackage{bm}

\usepackage{epstopdf}

\usepackage[T1]{fontenc}
\usepackage[applemac]{inputenc}
\usepackage{lmodern}
\usepackage[english]{babel}

\usepackage{ae}
\usepackage{units}

\usepackage{tikz}
\usetikzlibrary{arrows}

\usepackage{color}
\usepackage{url}

\usepackage[colorlinks]{hyperref}
\hypersetup{%
        plainpages=true,
        breaklinks=true,
        hypertexnames=false,
        pageanchor=true,
        colorlinks=true,
        linkcolor={blue},
        citecolor={red},
        urlcolor={blue},
        anchorcolor={black}
      }

\newcommand{\figref}[1]{\mbox{Fig.~\ref{#1}}}

\newcommand{\secref}[1]{\mbox{Sec.~\ref{#1}}}

\renewcommand{\eqref}[1]{\mbox{Eq.~(\ref{#1})}}

\newcommand{\ket}[1]{|#1\rangle}

\newcommand{\ketbra}[2]{\left| #1 \rangle \langle #2 \right|}
\newcommand{\brakket}[3]{\left\langle #1\left| #2 \right| #3\right\rangle}
\newcommand{\expec}[1]{\left\langle #1 \right\rangle}

\newcommand{\sx}{\hat \sigma_x}
\newcommand{\sm}{\hat \sigma_-}
\renewcommand{\sp}{\hat \sigma_+}

\newcommand{\abs}[1]{\left|#1\right|}

\newcommand{\abssq}[1]{\left| #1 \right|^2}

\newcommand{\nn}{\nonumber}

\newcommand{\be}{\begin{equation}}
\newcommand{\ee}{\end{equation}}
\newcommand{\bea}{\begin{eqnarray}}
\newcommand{\eea}{\end{eqnarray}}

\begin{document}

\title{Cutting Feynman Loops in Ultrastrong Cavity QED: Stimulated Emission and Reabsorption of Virtual Particles Dressing a Physical Excitation}

\author{Omar Di Stefano}
\affiliation{Dipartimento di Fisica e di Scienze della Terra, Universit\`{a} di Messina, I-98166 Messina, Italy}

\author{Roberto Stassi}
\affiliation{Dipartimento di Fisica e di Scienze della Terra, Universit\`{a} di Messina, I-98166 Messina, Italy}
\affiliation{CEMS, RIKEN, Saitama 351-0198, Japan}

\author{Luigi Garziano}
\affiliation{Dipartimento di Fisica e di Scienze della Terra, Universit\`{a} di Messina, I-98166 Messina, Italy}
\affiliation{CEMS, RIKEN, Saitama 351-0198, Japan}

\author{Anton Frisk Kockum}
\affiliation{CEMS, RIKEN, Saitama 351-0198, Japan}

\author{Salvatore Savasta}
\affiliation{Dipartimento di Fisica e di Scienze della Terra, Universit\`{a} di Messina, I-98166 Messina, Italy}
\affiliation{CEMS, RIKEN, Saitama 351-0198, Japan}

\author{Franco Nori}
\affiliation{CEMS, RIKEN, Saitama 351-0198, Japan}
\affiliation{Physics Department, The University of Michigan, Ann Arbor, Michigan 48109-1040, USA}

\date{\today}

\begin{abstract}

In quantum field theory, bare particles are dressed by a cloud of virtual particles to form physical particles. The virtual particles affect properties such as the mass and charge of the physical particles, and it is only these modified properties that can be measured in experiments, not the properties of the bare particles. The influence of virtual particles is prominent in the ultrastrong-coupling regime of cavity quantum electrodynamics (QED), which has recently been realized in several condensed-matter systems. In some of these systems, the effective interaction between atom-like transitions and the cavity photons can be switched on or off by external control pulses. This offers unprecedented possibilities for exploring quantum vacuum fluctuations and the relation between physical and bare particles. Here we show that, by applying external electromagnetic pulses of suitable amplitude and frequency, each virtual photon dressing a physical excitation in cavity-QED systems can be converted into a physical observable photon, and back again. In this way, the hidden relationship between the bare and the physical excitations becomes experimentally testable. The conversion between virtual and physical photons can be clearly pictured using Feynman diagrams with cut loops.

\end{abstract}


\maketitle

\section{Introduction}

In quantum field theory, the creation and annihilation operators in the Lagrangian describe the creation and destruction of \emph{bare} particles which, however, cannot be directly observed in experiments (see, \emph{e.g.}, Refs.~\cite{Peskin1995, Maggiore2004}). Bare particles, due to the interaction terms in the Lagrangian, are actually \emph{dressed} by \emph{virtual} particles and become real \emph{physical} particles which can be detected. The interaction modifies the properties of the particles, \emph{e.g.}, giving rise to the Lamb shift of electronic energy levels \cite{Lamb1947, Bethe1947} and affecting the charge, mass, and magnetic moment of the electron \cite{Schwinger1948, Welton1948, Peskin1995}. The predictions of the theory must be expressed in terms of the properties of the physical particles, not of the non-interacting (or bare) particles \cite{Peskin1995, Maggiore2004}. The relations between the bare and the physical particles, like the bare particles themselves, are unobservable. 

The influence of virtual particles features prominently in the ultrastrong coupling (USC) regime of cavity quantum electrodynamics (QED) \cite{Stassi2013, Garziano2014}. In cavity QED \cite{Haroche2013}, the interaction between light confined in a reflective cavity and natural or artificial atoms is studied in conditions where the quantum nature of light is important. The system enters the USC regime when the light-matter coupling rate becomes an appreciable fraction of the unperturbed resonance frequencies of the photons and the atom. In this regime, the routinely-invoked rotating wave approximation (RWA) is no longer applicable and the nonresonant terms in the light-matter interaction significantly change the standard cavity QED scenarios \cite{Dimer2007, DeLiberato2007, Cao2010, Cao2011, Ridolfo2011, Ridolfo2012, Ridolfo2013, Stassi2013, Sanchez-Burillo2014, Garziano2014, Cacciola2014, Lolli2015, Garziano2015, Garziano2016}. Recently, the USC regime has been reached experimentally in a variety of solid-state systems and spectral ranges \cite{Forn-Diaz2010, Niemczyk2010, Todorov2010, Schwartz2011, Scalari2012, Geiser2012, Kena-Cohen2013, Gambino2014, Maissen2014, Goryachev2014}.

The need to account for virtual particles in the USC regime of cavity QED is exemplified by the fact that the correct description of the output photon flux from the cavity, as well as of higher-order Glauber normal-order correlation functions, requires a proper generalization of input-output theory \cite{Ridolfo2012}. Due to the contribution from counter-rotating terms in the interaction Hamiltonian, the ground state $\ket{E_0}$ of the system contains a finite number of photons \cite{Ashhab2010}, \emph{i.e.}, $\brakket{E_0}{\hat a^\dag \hat a}{E_0} \neq 0$, where $\hat a$ and $\hat a^\dag$ are the annihilation and creation operators for the cavity mode. However, the ground state cannot emit energy, so the output photon flux cannot be proportional to $\expec{\hat a^\dag \hat a}$, as in standard input-output theory. Instead, it has been shown \cite{Ridolfo2012, Garziano2013} that the cavity output (which can be detected by a photo-absorber) is proportional to $\expec{\hat x^- \hat x^+}$, where $\hat x^+$ is the positive frequency component of the quadrature operator $\hat x = \hat a + \hat a^\dag$ and $\hat x^- = (\hat x^+)^\dag$. Since $\brakket{E_0}{\hat x^- \hat x^+}{E_0} = 0$, the photons that contribute to the ground state are not physical observable particles, but virtual ones. Furthermore, also the (physical) system excitations are enriched by unobservable virtual particles. For instance, the first excited state, corresponding to a single physical particle, may contain contributions from an odd number of virtual particles. All these virtual contributions, however, are significant only in the USC regime, not at weaker coupling strengths.

In QED, the modifications of the electron properties due to the interaction with virtual particles are known as radiative corrections \cite{Peskin1995}. In addition to the diagrams describing the processes in lowest order of perturbation theory, the Feynman diagrams representing the radiative corrections to a process contain additional vertices (loop diagrams), corresponding to the emission and re-absorption of virtual photons. Here we show that, analogously, the energy corrections to the ground state and to the excited states of a cavity-QED system in the USC regime are described by loop diagrams corresponding to the emission and re-absorption of virtual photons. 

An interesting feature of these condensed-matter systems is that the effective interaction between atom-like transitions and the cavity field can be switched on and off by applying external drives. This offers the opportunity to convert the virtual excitations into real particles which can be detected. Both spontaneous \cite{Stassi2013} and stimulated \cite{Huang2014} conversion of virtual photons from the ground state of a cavity QED system in the USC regime have recently been analyzed . Also, virtual photon pairs are converted into real ones in the dynamical Casimir effect (DCE) \cite{Moore1970}, which has been analyzed for \cite{Johansson2009, Johansson2010, Johansson2013} and experimentally demonstrated \cite{Wilson2011} in circuit QED. Potentially, a proper modulation of the mirror in a DCE setup could also allow for absorption of photon pairs \cite{DeSousa2015}. In contrast to these previous works, we here show how to convert various numbers of virtual photons into real ones and back, both for the dressed vacuum state and for a dressed excited states. We show that the corresponding Feynman diagrams can be obtained by cutting the loop diagrams describing the energy correction of a physical excitation. Specifically, conversion of virtual photons dressing a physical excitation into real ones is described by the first half of cut loop-diagrams (photon emission). Similarly, the conversion of real photons back into virtual ones bound to a physical excitation corresponds to the second half (photon absorption). Moreover, the proposed scheme, does not need ultrafast modulation of boundary conditions and it can give rise to a conversion probability close to one.

\section{Results}

\subsection{The Rabi model}

The simplest cavity-QED model beyond the RWA is the quantum Rabi model \cite{Rabi1937, Braak2011}. The Hamiltonian is ($\hbar =1$) $\hat H_{\rm R} =\hat H_0 +\hat V$, where $\hat H_0 = \omega_{\rm c}\, \hat a^\dag \hat a  + \omega_{e} \ketbra{e}{e} + \omega_g \ketbra{g}{g}$ is the bare Hamiltonian in the absence of interaction. Here, $\hat a$ and $\hat a^\dag$ are the photon destruction and creation operators for the cavity mode with resonance frequency $\omega_{\rm c}$, $\ket{g}$ and $\ket{e}$ are the ground and excited atomic states, respectively, and $\omega_{e(g)}$ are the corresponding energy eigenvalues. The interaction Hamiltonian is 
\begin{equation}
\hat V = \Omega_{\rm R} \left(\hat a^\dag +\hat a \right) \sx,
\label{VR} 
\end{equation}
where $\Omega_{\rm R}$ is the coupling strength and $\sx = \sp + \sm = \ketbra{e}{g} + \ketbra{g}{e}$. When $\omega_{\rm c} \approx \omega_{eg} \equiv \omega_e - \omega_g$, the interaction Hamiltonian can be separated into a resonant and a nonresonant part: $\hat V = \hat V_{\rm r} + \hat V_{\rm nr}$, where $\hat V_{\rm r} = \Omega_{\rm R} (\hat a^\dag \sm + \hat a \sp)$, and $\hat V_{\rm nr} = \Omega_{\rm R}( \hat a^\dag \sp + \hat a \sm)$. The nonresonant terms do not conserve energy nor the number of excitations. They can be neglected when $\Omega_{\rm R}/(\omega_{\rm c} + \omega_{eg}) \ll 1$.

\begin{figure}
	\includegraphics[width=120 mm]{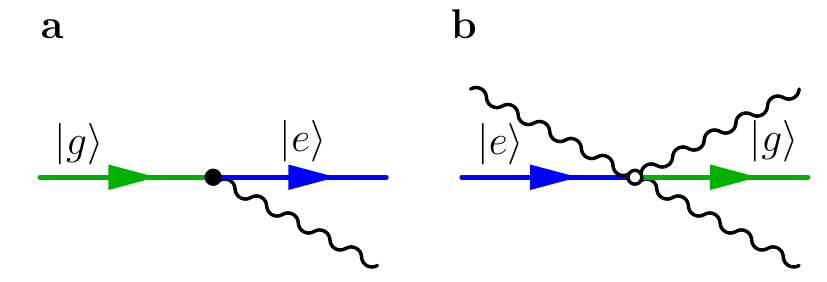}
\caption{Diagrams for processes in the Rabi model. The horizontal lines, coloured blue for $\ket{e}$ and green for $\ket{g}$, represent the qubit states and the wavy lines the cavity photons. (a) Diagram for the transition $\ket{g,0} \to \ket{e,1}$, induced by the nonresonant term $\hat a^\dag \sp$. The filled vertex is used to mark nonresonant processes. (b) Diagram for the transition $\ket{e,1} \to \ket{g,2}$. This is a resonant process, induced by the term $\hat a^\dag \sm$, marked by an empty vertex. For a process with stimulated emission, such as this one, each photon involved is represented by a separate wavy line. This is the convention used in the rest of this article. \label{fig:Diagrams0GE1G2}}
\end{figure}

The interaction Hamiltonian has a structure which is very similar to that of the QED interaction potential, although it is less complicated. The Rabi model can be viewed as a very simple QED system, where there is only a single photon mode and a two-state electron. As a consequence, we would expect that Feynman diagrams for the Rabi Hamiltonian will be a simplified version of QED diagrams. One such diagram, for the nonresonant transition $\ket{g,0} \to \ket{e,1}$ (the second entry in the ket denotes the photon number), is shown in \figref{fig:Diagrams0GE1G2}a.

However, some care must be taken when drawing diagrams for processes involving more than one photon in the same mode \cite{Cohen-Tannoudji1998}, which occur in cavity QED. Stimulated emission \cite{Einstein1917}, the mechanism behind laser action, is one such process. It is a one-photon process $\ket{e,n} \to \ket{g,n+1}$, where the $n$ photons in the initial state stimulate the downward transition of the atom, affecting the transition rate which becomes proportional to $n+1$. This factor must be included in the rules for the diagrams in order for calculations to be correct. An example of a diagram showing the stimulated-emission process $\ket{e,1} \to \ket{g,2}$ is presented in \figref{fig:Diagrams0GE1G2}b. A more detailed discussion about diagrams for stimulated emission can be found in the Supplementary Material.

\subsection{Bare \emph{vs} physical excitations}

Owing to the presence of $\hat V_{\rm nr}$ in the Rabi Hamiltonian, the operator describing the total number of excitations, $\hat N = \hat a^\dag \hat a + \ketbra{e}{e}$, does not commute with $\hat H_{\rm R}$ and as a consequence the eigenstates of $\hat H_{\rm R}$ do not have a definite number of excitations \cite{Ashhab2010}. When $\hat V_{\rm nr}$ can be neglected, the Hamiltonian becomes block-diagonal and easy to diagonalize (this is the Jaynes-Cummings (JC) model \cite{Jaynes1963}). The resulting eigenstates can be labelled according to their definite number of excitations $n$. The ground state (zero excitations) is simply $\ket{\mathcal{E}_0} = \ket{g, 0}$, and the $n \geq 1$ excitation states $\ket{{\cal E}^\pm_n}$, obtained by diagonalization of $2 \times 2$ subspaces, can be written as
\begin{equation}
\ket{{\cal E}^+_n} = {\cal C}_{n}\ket{g,n}  + {\cal S}_{n} \ket{e,n-1} \qquad \ket{{\cal E}^-_n} = - {\cal S}_{n} \ket{g,n} +{\cal C}_{n} \ket{e,n-1},
\label{JC}
\end{equation}
where ${\cal C}_{n}$ and ${\cal S}_{n}$ are amplitudes determined by $\Omega_{\rm R}$ and the detuning $\omega_{\rm c} - \omega_{eg}$. The eigenstates $\ket{E_i}$ of the full Rabi Hamiltonian, however, are expressed as a superposition of bare states with varying numbers of bare excitations (see, \emph{e.g.}, Ref.~\cite{Garziano2013}):
\begin{equation}
\ket{E_i} = \sum_{k= 0}^\infty \left(c^i_{g, k} \ket{g, k} + d^i_{e,k} \ket{e,k} \right),
\label{zerotilde}
\end{equation}
where the coefficients $c^i_{g,k}$ and $d^i_{e,k}$ are determined by $\Omega_{\rm R}$, $\omega_{\rm c}$ and $\omega_{eg}$. When $\Omega_{\rm R} \ll \omega_{\rm c}, \omega_{eg}$, the Rabi eigenstates reduce to the JC ones. Note that while $\hat N$ is not conserved with the Rabi Hamiltonian, the parity (even or odd number of excitations) still is.

The mean photon number for the system in the ground state is 
\be
\brakket{E_0}{\hat a^\dag \hat a}{E_0} = \sum_k \left(2k \abssq{c^0_{g,2k}} +(2k+1) \abssq{d^0_{e, 2k+1}} \right).
\ee
However, these ground-state photons are virtual and cannot be detected. Otherwise the system, emitting a continuous stream of photons from its ground state, would be a perpetual-motion machine. A proper treatment shows that the output emission rate from a single-mode resonator is not proportional to $\expec{\hat a^\dag \hat a}$, but to $\expec{\hat x^- \hat x^+}$, where $\hat x^+$ is the positive frequency component of the quadrature operator $\hat x = \hat a + \hat a^\dag$ and $\hat x^- = (\hat x^+)^\dag$ \cite{Ridolfo2012, Garziano2013}. For weak coupling, $\expec{\hat a^\dag \hat a}$ and $\expec{\hat x^- \hat x^+}$ coincide, but in the USC regime they can differ markedly. 

The components $\hat x^+$ and $\hat x^-$ are obtained in the eigenvector basis of $\hat H_{\rm R}$ as $\hat x^+=\sum_{i<j}x_{ij} \ketbra{E_i}{E_j}$, where $x_{ij} = \brakket{E_i}{\hat x}{E_j}$, if the eigenstates of $\hat H_{\rm R}$ are labelled according to their eigenvalues such that $E_k > E_j$ for $k>j$. As expected, we find that  $\brakket{E_0}{\hat x^-(t) \hat x^+(t)}{E_0} = 0$, which demonstrates that the photonic Fock states enriching the Rabi ground state are actually virtual. This reasoning can be generalized to the excited states of the system. For the first excited state, the one-photon correlation is different from zero ($\brakket{E_1}{\hat x^-(t) \hat x^+(t)}{E_1} \neq 0$). However, the output coincidence rate from this state, proportional to the \emph{physical} two-photon correlation function $\brakket{E_1}{(\hat x^-)^2 (\hat x^+)^2}{E_1}$, is equal to zero. On the contrary, the correlation functions for $n \geq 2$ bare photons in the first excited state are different from zero ({\it e.g.}, $\brakket{E_1}{(\hat a^\dag)^2 (\hat a)^2}{E_1} \neq 0$).  We can conclude that $\ket{E_{1}}$ is a single physical excitation which, however, is enriched by a larger number of virtual photons.

\subsection{Energy corrections and loop diagrams}

The analytical spectrum of $H_{\rm R}$ is defined in terms of the power series of a transcendental function \cite{Braak2011}. Approximate forms, which may provide more insight, can be derived by a perturbative approach (see, \emph{e.g.}, Ref.~\cite{Konishi2009}). Let us consider the correction to the ground state energy $\Delta_0 \equiv E_0 - {\cal E}_0$. The lowest-order (in the nonresonant potential) contribution can be expressed as
\begin{equation}
\Delta^{(2)}_0 = \brakket{g,0}{\hat V_{\rm nr} \hat G({\cal E}_{0}) \hat V_{\rm nr} }{g,0},
\label{delta0}
\end{equation}
where $\hat G(z) = (z - \hat H_0 + \hat V_{\rm r})^{-1}$ is the JC Green's function. The Green's function $\hat G(z)$ can be directly calculated by using the JC eigenstates from \eqref{JC}. Alternatively, it can be expressed in a Dyson series containing $\hat V_{\rm r}$ and the Green's function $\hat G_0(z) = (z - \hat H_0)^{-1}$ in the absence of interaction: $\hat G = \hat G_0 + \hat G_0 \hat V_{\rm r} \hat G_0 + \hat G_0 \hat V_{\rm r} \hat G_0 \hat V_{\rm r} \hat G_0+ \dots$. Equation (\ref{delta0}) can thus be expanded as
\begin{equation}
\Delta^{(2)}_0 = \brakket{g, 0}{\hat V_{\rm nr} \hat G_0({\cal E}_{0}) \hat V_{\rm nr}}{g, 0} + \brakket{g, 0}{\hat V_{\rm nr} \hat G_0({\cal E}_{0}) \hat V_{\rm r} \hat G_0({\cal E}_{0}) \hat V_{\rm nr}}{g, 0} + \ldots
\label{delta0exp}
\end{equation}
A direct inspection of the terms in the series shows that only the terms with an even number of $V_{\rm r}$ are different from zero. It is possible to associate a diagram with each of the terms in the series appearing in \eqref{delta0exp}.
\begin{figure}
\includegraphics[width=140 mm]{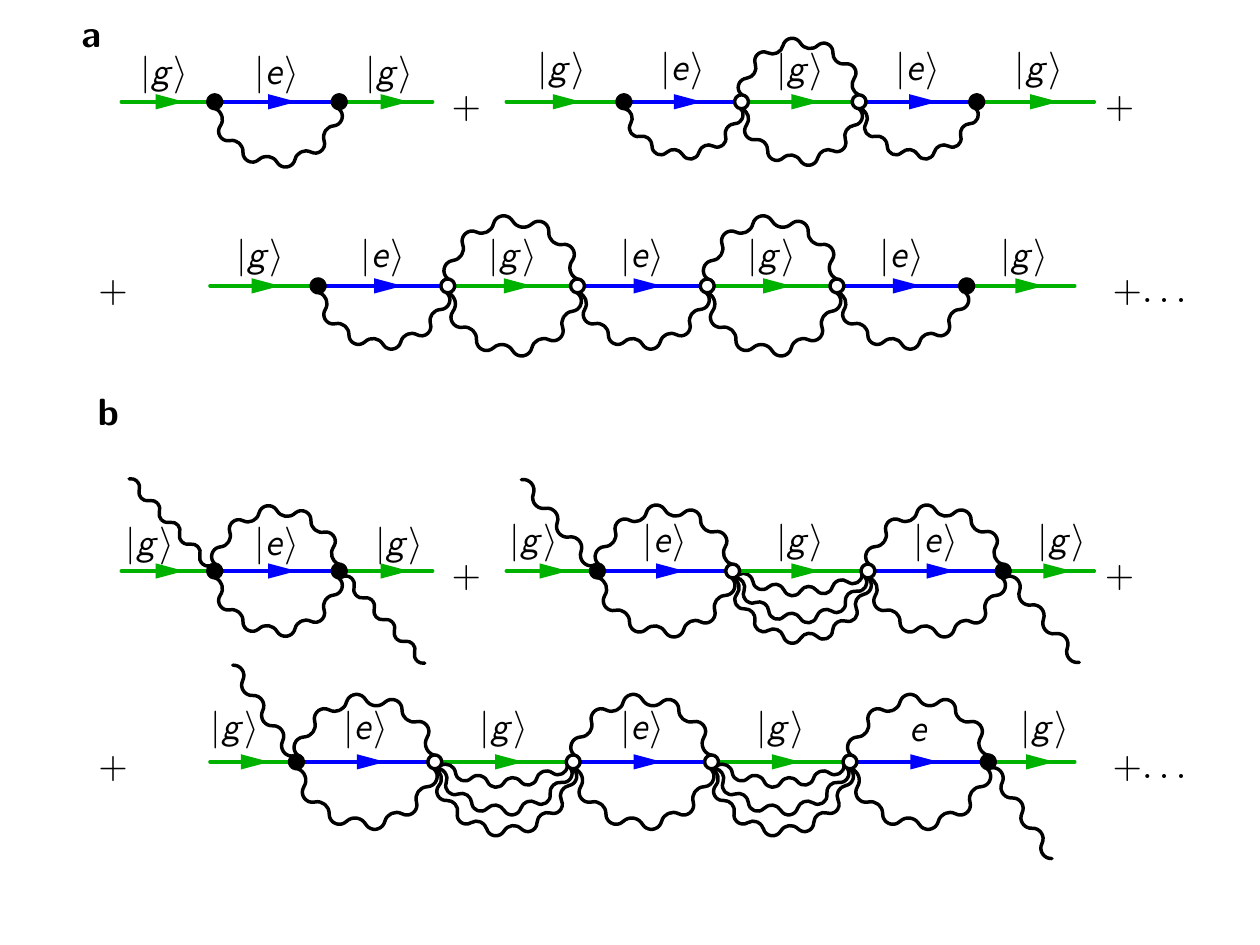}
\caption{Feynman diagrams contributing to the energy correction of the ground state (a), and of the first excited state (b) of the Rabi Hamiltonian. Each bubble diagram, corresponding to a matrix element of $\hat G_0$, describes the intermediate virtual excitations enriching the ground and the first excited states. The virtual excitations originate from the nonresonant terms in the interaction Hamiltonian. \label{fig:FeynmanFirst3Diagrams}}
\end{figure}
Figure \ref{fig:FeynmanFirst3Diagrams}a shows the first three diagrams providing a nonzero contribution. The first corresponds to the first term in the r.h.s.~of  \eqref{delta0exp}. The second diagram describes the third term in the series: $\brakket{g, 0}{\hat V_{\rm nr} \hat G_0({\cal E}_{0}) \hat V_{\rm r} \hat G_0({\cal E}_{0}) \hat V_{\rm r} \hat G_0({\cal E}_{0}) \hat V_{\rm nr}}{g, 0}$. Each bubble diagram, corresponding to a matrix element of $\hat G_0$, describes intermediate virtual excitations. All the resulting bubble diagrams contain at most two photon waves, since we considered only the lowest-order corrections in the nonresonant potential. Four and more photon waves arise when going beyond second-order perturbation theory.

This approach can also be applied to the excited states. Considering the first excited state, we obtain
\begin{equation}
\Delta^{(2)}_1 = \brakket{{\cal E}^+_1}{\hat V_{\rm nr} \hat G({\cal E}^+_1) \hat V_{\rm nr}}{{\cal E}^+_1} = {\cal C}^2_1 \brakket{g,1}{\hat V_{\rm nr} \hat G({\cal E}^+_1) \hat V_{\rm nr}}{g, 1}.
\label{delta1}
\end{equation}
The mean value over the state $\ket{g,1}$ in \eqref{delta1} can be expanded by exploiting the Dyson series. The corresponding first three diagrams providing a nonzero contribution are displayed in \figref{fig:FeynmanFirst3Diagrams}b. In the diagrams we find internal loops where two or three virtual photons are created and finally reabsorbed. The initial and final physical states, however, only contain a one-photon wave. The energy corrections $\Delta^{(2)}_0$ and $\Delta^{(2)}_1$ can be easily evaluated by directly using $\hat G(z)$ or by summing up the infinite contributions arising from the Dyson series and described by the diagrams. These calculations, and a comparison between the approximate analytical energy corrections and the corresponding nonperturbative numerical calculations, can be found in the Supplementary Material.

\subsection{Three-level atom}

\begin{figure}
\includegraphics[width=90 mm]{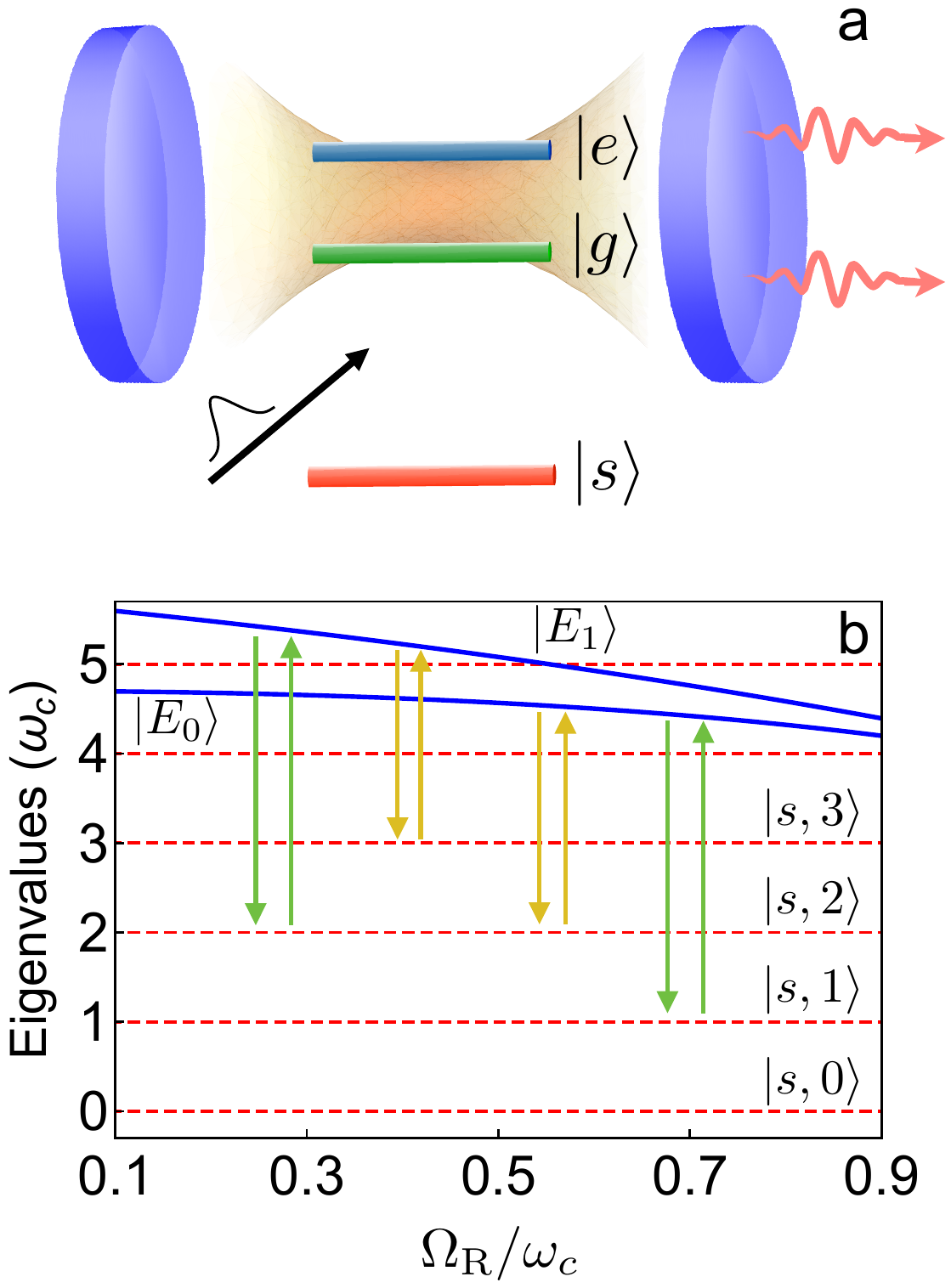}
\caption{(a) Schematic of the system in which a three-level atom is placed in a cavity.  The upper two levels $\ket{e}$ and $\ket{g}$ of the atom resonantly couple to a single cavity mode. The effective atom-cavity interaction can be controlled by external electromagnetic pulses (arrow with a gaussian pulse) inducing transitions from the cavity-interacting levels $\ket{g}$ and $\ket{e}$ to the noninteracting level $\ket{s}$ and \emph{vice versa}. These pulses can induce the emission of photons (red arrows) enriching the ground or the excited states of the Rabi Hamiltonian. (b) Lowest energy-levels of the system as a function of the normalized coupling strength $\Omega_{\rm R}/ \omega_{\rm c}$ and the transitions stimulated by the external pulses. Yellow arrows mark transitions induced by $\hat V_{sg}$ and green arrows mark transitions induced by $\hat V_{se}$. \label{fig:SchematicAndEnergyLevels}} 
\end{figure}

We now consider a system consisting of a single-mode cavity interacting with the upper two levels $\ket{e}$ and $\ket{g}$ of a three-level atom. The energy difference $E_{gs}$ between the bottom level $\ket{s}$ and the middle level $\ket{g}$ is assumed to be much larger than the cavity-mode resonance frequency such that the cavity does not interact with the atom in the lowest energy state $\ket{s}$ (see \figref{fig:SchematicAndEnergyLevels}a). As we will show, the additional state $\ket{s}$ enables an effective on/off-switch of the atom-cavity interaction. The system Hamiltonian is simply $\hat H_{\rm C} = \hat H_{\rm R} + \omega_s \ketbra{s}{s}$. This Hamiltonian is block-diagonal and its eigenstates can be separated into a non-interacting sector $\ket{s,n}$, with energy $\omega_{s}+ n \omega_{\rm c}$, where $n$ labels the cavity photon number, and dressed atom-cavity states $\ket{E_i}$, resulting from the diagonalization of the Rabi Hamiltonian (see \figref{fig:SchematicAndEnergyLevels}b).

The direct excitation of the atom by applied electromagnetic pulses is described by the Hamiltonian
\begin{equation}
\hat H_{\rm d} = {\cal E}_{\rm d}(t)(\hat V_{sg} +\hat V_{se}),
\label{Hd}
\end{equation}
where $\hat V_{sg} = \mu_{sg} (\ketbra{g}{s} + \ketbra{s}{g})$, $\hat V_{se} = \mu_{se} (\ketbra{e}{s} + \ketbra{s}{e})$, and $\mu_{sg}$ and $\mu_{se}$ are the dipole moments (here assumed to be real) for the transitions $\ket{s} \leftrightarrow \ket{g}$ and $\ket{s} \leftrightarrow \ket{e}$, respectively. We consider quasi-monochromatic pulses ${\cal E}_{\rm d}(t) = A(t) \cos(\omega t)$, where ${A}(t)$ is a Gaussian envelope. We only consider pulses which are out of resonance with the transition $\ket{g} \leftrightarrow \ket{e}$ and neglect this transition in \eqref{Hd}. If the system is prepared in a dressed state $\ket{E_i}$, the driving Hamiltonian $\hat H_{\rm d}$ can induce transitions towards the noninteracting states $\ket{s,m}$:
\begin{equation}
\hat H_{\rm d} \ket{E_i} = {\cal E}_{\rm d}(t) \sum_{k= 0}^\infty \left( \mu_{sg} c^i_{g,k} \ket{s,k} + \mu_{se} d^i_{e,k} \ket{s,k} \right).
\label{eq}
\end{equation}
Thus $\hat H_{\rm d}$ applied to a dressed state is able to convert the virtual photons enriching the physical excitations into real ones which can be detected. This is possible because $\hat H_{\rm d}$ induces transitions from the atomic states $\ket{g}$ and  $\ket{e}$ (coupled  to the cavity) to the noninteracting state $\ket{s}$. Of course, the transitions only occur if the driving-field frequency $\omega$ is resonant with the corresponding transition frequency. In the absence of counter-rotating terms, a JC eigenstate with $n$ excitations can only undergo transitions towards states with $n$ photons: $\ket{E_n^\pm} \to \ket{s,n}$ (for $\mu_{sg} \neq 0$), or $n-1$ photons: $\ket{E_n^\pm} \to \ket{s,n-1}$ (for $\mu_{se} \neq 0$).

\subsection{Stimulated emission and reabsorption of virtual particles}

We first consider the system prepared in the ground state $\ket{E_0}$ of the Rabi Hamiltonian. An input pulse of central frequency $\omega \simeq E_0 - \omega_s - 2 \omega_{\rm c}$ can induce a transition $\ket{E_0} \to \ket{s,2}$, corresponding to a stimulated emission process (see \figref{fig:SchematicAndEnergyLevels}b). The corresponding matrix element $\brakket{s, 2}{\hat V_{sg}}{E_0} = \mu_{sg} c^0_{g,2}$, determining the transition probability, is proportional to the probability amplitude $c^0_{g,2}$ that in the Rabi ground state there are two virtual photons. By exploiting second-order perturbation theory, this matrix element can be expressed as $\brakket{s, 2}{\hat V_{sg} \hat G({\cal E}_0) \hat V_{\rm nr}}{E_0}$. From the Dyson series, we obtain
\begin{eqnarray}
\brakket{s, 2}{\hat V_{sg} \hat G({\cal E}_0) \hat V_{\rm nr}}{E_0} &=& \brakket{s,2}{\hat V_{\rm sg} \hat G_0({\cal E}_0) \hat V_{\rm nr}}{E_0} + \brakket{s, 2}{\hat V_{sg} \hat G_0({\cal E}_0) \hat V_{\rm r} \hat G_0({\cal E}_0) \hat V_{\rm nr}}{E_0} \nn\\ 
&&+ \brakket{s, 2}{ \hat V_{sg} \hat G_0({\cal E}_0) \hat V_{\rm r} \hat G_0({\cal E}_0) \hat V_{\rm r} \hat G_0({\cal E}_0) \hat V_{\rm nr} }{E_0} + \ldots
\label{s2}
\end{eqnarray}
%
\begin{figure}
\includegraphics[width=130 mm]{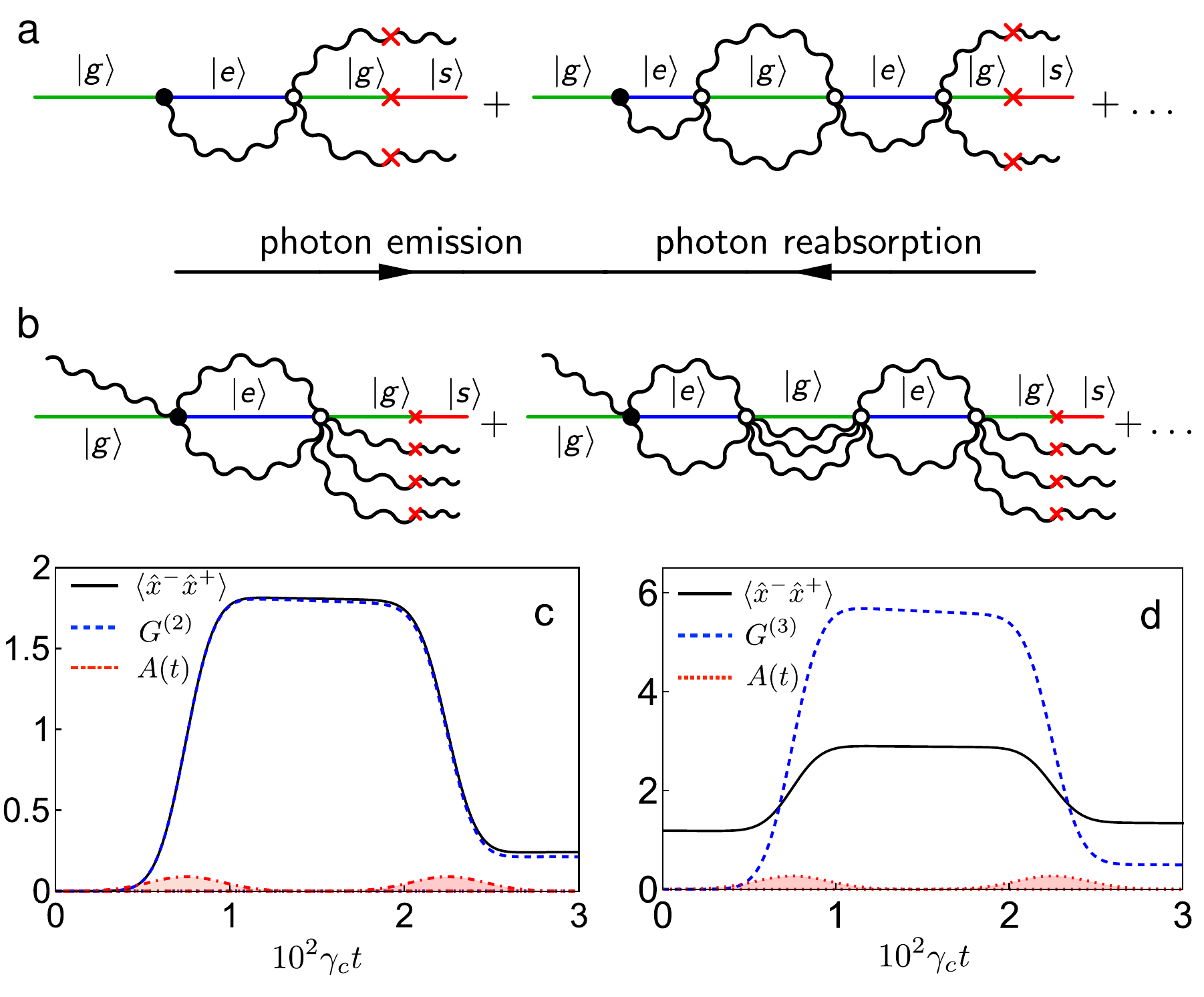}
\caption{(a) Diagrams contributing to the transition matrix element $\brakket{s,2}{\hat V_{sg}}{E_0}$, associated with the transition $\ket{E_0} \to \ket{s,2}$ (rightward time-arrow) where two cavity photons are emitted. The same diagrams but with a leftward time-arrow describe the reverse transition $\ket{s,2} \to \ket{E_0}$, where two cavity photons get trapped into the Rabi ground state. (b) Diagrams contributing to the matrix element $\brakket{s,3}{\hat V_{sg}}{E_1}$, associated with the transitions $\ket{E_1} \leftrightarrow \ket{s,3} $, where three photons enriching the lowest-energy excited state of the Rabi Hamiltonian are emitted or reabsorbed back. The red crosses represent the perturbation $\hat V_{sg}$. (c) Numerical calculations of the mean cavity-photon number (continuous black curve), and of the two-photon correlation function (dashed blue curve) corresponding to diagrams in \figref{fig:Vsg}a . The system is initially prepared in the state $\ket{E_0}$. A $\pi$ pulse, resonantly exciting the system from $\ket{E_0}$ to $\ket{s,2}$, is then sent. A second (red) pulse induces the  transition back from $\ket{s,2}$ to $\ket{E_0}$. (d) Numerical calculations of the mean cavity-photon number (continuous black curve), and of the two-photon correlation function (dashed blue curve) corresponding to diagrams in \figref{fig:Vsg}b. The system is initially prepared in the state $\ket{E_1}$. A $\pi$ pulse (filled curve), resonantly exciting the system from $\ket{E_1}$ to $\ket{s,2}$, is then sent. A second pulse induces the transition back from $\ket{s,2}$ to $\ket{E_1}$. Here we used $\Omega_{\rm R}/\omega_{\rm c} = 0.15$, $\omega_{eg}=\omega_{\rm c}$, and $\gamma_{eg} = \gamma_{gs} = \gamma_{c} = 2\times 10^{-5}\omega_{\rm c}$, where $\gamma_c$ is the decay rate for the cavity photons and $\gamma_{eg}, \gamma_{gs}$ the decay rates for the atom transitions $\ket{e} \to \ket{g}$ and $\ket{g} \to \ket{s}$. \label{fig:Vsg}}
\end{figure}

Figure \ref{fig:Vsg}a displays the diagrams describing the first nonzero terms in this series. The red crosses represent the action of the perturbation $\hat V_{sg}$. These Feynman diagrams provide a clear interpretation of the emission process. The loops in \figref{fig:FeynmanFirst3Diagrams} contain virtual photons which contribute to the energy correction of the state $\ket{E_0}$ and $\ket{E_1}$. As shown in \figref{fig:Vsg}, the time-dependent perturbation $\hat V_{sg}$ is able to cut these loops. These diagrams show that the virtual photons in the loops are not just a technical feature of perturbation theory but describes internal physical processes, which can be interrupted by a suitable perturbation able to convert each virtual photon into an observable physical photon. Specifically, diagrams in \figref{fig:Vsg}a together with the rightward time-arrow describe the transition $\ket{E_0} \to \ket{s,2}$, where two cavity photons are emitted. The same diagrams, but with a leftward time-arrow, describe the transition $\ket{s,2} \to \ket{E_0}$, where two cavity photons are reabsorbed into the Rabi ground state. The potential $\hat V_{sg}$ induces the breaking of two-photon loops, converting virtual photon pairs into real ones. It is not able, however, to break one-photon loops. These can be broken by the potential $\hat V_{se}$ as shown below.

It is even more interesting to undress the excited states of the Rabi model. This can provide access to the relationship between bare and physical excitations. Let us consider the lowest-energy excited state $\ket{E_1}$ which, as we have shown above, is a single-particle state. Following the same steps as used in obtaining the series in \eqref{s2}, the diagrams in \figref{fig:Vsg}b can be drawn. According to the Fermi golden rule, an input pulse of central frequency $\omega \simeq E_1 - \omega_s - 3 \omega_{\rm c}$ can induce a transition $\ket{E_1} \to \ket{s, 3}$. The corresponding matrix element $\brakket{s, 3}{\hat V_{sg}}{E_1} = \mu_{sg} c^1_{g,3}$ is proportional to the probability amplitude that in the state $\ket{E_1}$ there are three virtual photons. By applying second-order perturbation theory, it can be expressed as
\begin{equation}
\brakket{s, 3}{\hat V_{sg}}{E_1} ={\cal C}_1 \brakket{s, 3}{ \hat V_{sg} \hat G({\cal E}^-_1) \hat V_{\rm nr} }{g,1}.
\end{equation}
The analytical perturbative calculations of the matrix elements $\brakket{s, 2}{\hat V_{sg}}{E_0}$ and $\brakket{s, 3}{\hat V_{sg}}{E_1}$ are described in the Supplementary Material. 

We complete the above analysis by presenting nonperturbative numerical calculations which accurately describe the dynamics of the undressing and re-dressing of the Rabi vacuum and of the Rabi lowest-energy excitation. We take into account the presence of dissipation channels, the presence of higher energy levels, and the non-monochromaticity of the driving pulses.  All the dynamical evolutions displayed in Figs.~\ref{fig:Vsg}c and \ref{fig:Vsg}d have been calculated numerically solving the master equation $\dot{\hat \rho}(t) = i [\hat \rho(t), \hat H_{\rm C}] + \sum_{j} \mathcal{\hat L}_{j}\hat \rho(t)$ \cite{Breuer2002, Beaudoin2011, Garziano2013}, where $\mathcal{\hat L}_{j}$ are Liouvillian superoperators describing the different (atomic and photonic) dissipation channels. All calculations have been carried out with zero-temperature reservoirs.

We consider the system initially prepared in the state $\ket{E_0}$ (preparation starting from the ground state $\ket{s,0}$ can be easily achieved by sending a suitable $\pi$ pulse). Then, a Gaussian pulse with central frequency $\omega = E_0 - \omega_s - 2 \omega_{\rm c}$ induces the transition $\ket{E_0} \to \ket{s,2}$. Specifically, the pulse area required to obtain a complete transition is $\pi/ \abs{\brakket{s, 2}{\hat V_{sg}}{E_0}}$. The pulse arrival-time corresponds to the time when the loops in the Feynman diagrams are cut. Figure \ref{fig:Vsg}c displays the dynamics of the intracavity mean excitation number $\expec{\hat x^- \hat x^+}$, which is directly related to the output photon flux $\Phi_{\rm out}(t) = \gamma_c \expec{\hat x^-(t) \hat x^+(t)}$ (where $\gamma_c$ is the photon escape rate through the cavity boundary), as well as the equal-time second-order correlation function $G^{(2)}(t) = \expec{(\hat x^-(t))^2 (\hat x^+(t))^2}$. Before the arrival of the Gaussian pulse (shaded red curve), the output photon flux is zero, since $\brakket{E_0}{\hat x^- \hat x^+}{E_0} = 0$. After the arrival of the pulse, the photon flux becomes nonzero and $G^{(2)}(t) \simeq \expec{\hat x^- \hat x^+}$, confirming that a two-photon state is actually generated as expected from the diagrams in \figref{fig:Vsg}a. When a second pulse is sent, the two photons are reabsorbed into the Rabi ground state: $\ket{s, 2} \to \ket{E_0}$ (diagrams in \figref{fig:Vsg}a with the leftward time-arrow). Figure \ref{fig:Vsg}c shows that a residual small excitation remains in the system after the arrival of the second pulse. This can be attributed to the influence of damping and to a non-negligible transition probability to higher-energy levels induced by the tails of the pulse spectrum.

Figure \ref{fig:Vsg}d displays the dynamics starting from the system prepared in the state $\ket{E_1}$. We observe that, after the arrival of the Gaussian pulse (with central frequency $\omega = E_1 - \omega_s - 3 \omega_{\rm c}$ and area $\pi/ \abs{\brakket{s, 3}{\hat V_{sg}}{E_1}}$), the initial zero third-order correlation function $G^{(3)}$ approaches 6, the value corresponding to a three-photon state. This result confirms the occurrence of the transition $\ket{E_1} \to \ket{s, 3}$. Also in this case, the emitted photons are reabsorbed by sending an additional identical Gaussian pulse. We observe that, within the standard RWA, $\brakket{E_1}{\hat x^-(t) \hat x^+(t)}{E_1} = 1$. Figure \ref{fig:Vsg}d at $t=0$ displays a higher value. This is a peculiar effect of the USC regime, where the intracavity mean excitation number is quadrature-dependent. In particular, it increases for $\hat x$ measurements and decreases for measurements of the conjugate quadrature $\hat y = i (\hat a^\dag - \hat a)$.

\begin{figure}[!ht]
\includegraphics[width=130 mm]{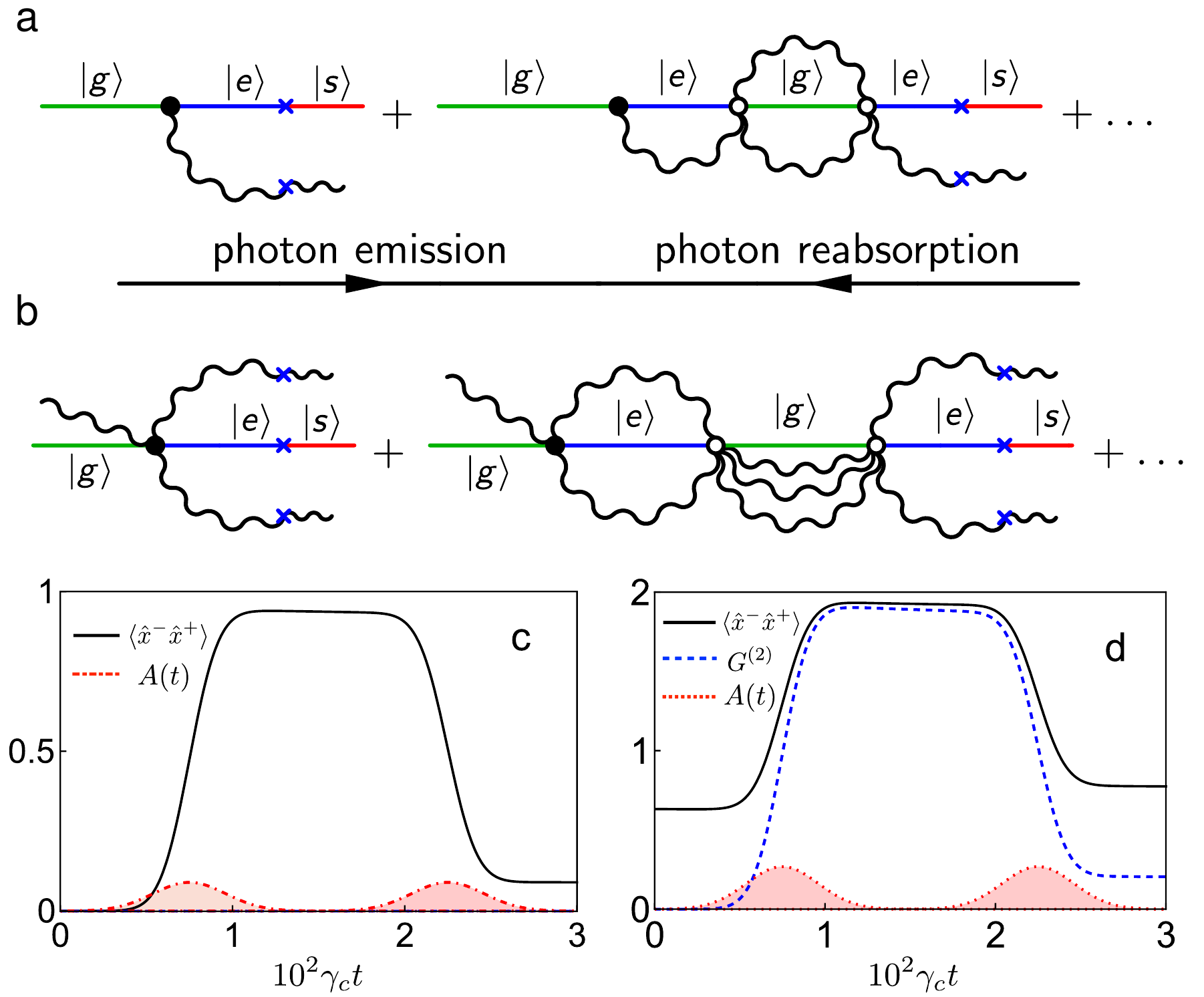}
\caption{(a) Diagrams contributing to the matrix element $\brakket{s,1}{\hat V_{se}}{E_0}$, associated with the transition $\ket{E_0} \to \ket{s,1}$ (rightward time-arrow) where a cavity photon is emitted. The same diagrams but with a leftward time-arrow describe the reverse transition $\ket{s,1} \to \ket{E_0}$, where a cavity photon is absorbed into the Rabi ground state.
(b) Diagrams contributing to the matrix element $\brakket{s,2}{\hat V_{se}}{E_1}$, associated with the transitions $\ket{E_1} \leftrightarrow \ket{s,2}$, where two photons enriching the lowest-energy excited state $\ket{E_1}$ of the Rabi Hamiltonian are emitted or reabsorbed back. The blue crosses represent the perturbation $\hat V_{se}$.
(c) Numerical calculations of the mean cavity-photon number (continuous black curve) corresponding to diagrams in \figref{fig:Vse}a. The system is initially prepared in the state $\ket{E_0}$. A $\pi$ pulse (shown in red), resonantly exciting the system from $\ket{E_0}$ to $\ket{s,1}$, is then sent. A second pulse induces the transition back from $\ket{s,1}$ to $\ket{E_0}$.
(d) Numerical calculations of the mean cavity-photon number (continuous black curve), and of the three-photon correlation function (dashed blue curve) corresponding to diagrams in \figref{fig:Vse}b. The system is initially prepared in the state $\ket{E_{1}}$. A $\pi$ pulse (red filled curve), resonantly exciting the system from $\ket{E_1}$ to $\ket{s,2}$, is then sent. A second pulse induces the transition back from $\ket{s,2}$ to $\ket{E_1}$. For the other parameters of the simulation not specified here, the same values as in \figref{fig:Vsg} were used. \label{fig:Vse}}
\end{figure}

Having studied above processes with $\hat V_{se}$, we now turn to those involving $\hat V_{se}$ instead. Figure~\ref{fig:Vse} shows these processes, with the action of $\hat V_{se}$ represented in the diagrams by blue crosses. These processes are able to break one-photon loops, as illustrated in \figref{fig:Vse}a, which shows the diagrams associated with the transition $\ket{E_0} \to \ket{s,1}$, where a cavity photon is emitted (rightward time-arrow) and reabsorbed (leftward time-arrow). Figure~\ref{fig:Vse}b shows the diagrams associated with the transitions $\ket{E_1} \leftrightarrow \ket{s,2}$, where two photons enriching the lowest-energy excited state $\ket{E_1}$ of the Rabi Hamiltonian are emitted or reabsorbed. The analytical perturbative calculations of the matrix elements $\brakket{s, 1}{\hat V_{se}}{E_0}$ and $\brakket{s, 2}{\hat V_{se}}{E_1}$ are described in the Supplementary Material. 

In complete analogy with what was shown in \figref{fig:Vsg}c and \figref{fig:Vsg}d, we present in \figref{fig:Vse}c and \figref{fig:Vse}d nonperturbative numerical calculations describing the dynamics of the undressing and re-dressing of the Rabi vacuum and of the Rabi lowest-energy excitation, taking into account the presence of dissipation channels, the presence of higher-energy levels, and the non-monochromaticity of the driving pulses. The dynamics of the intracavity mean excitation number $\expec{\hat x^- \hat x^+}$, which becomes close to $1$ shown in \figref{fig:Vse}c, and the equal-time second-order correlation function $G^{(2)}(t) \simeq \expec{\hat x^- \hat x^+}$, shown in \figref{fig:Vse}d, confirm that one-photon and two-photon states are actually generated as expected from the diagrams in \figref{fig:Vse}a and \figref{fig:Vse}b, respectively. We observe that both in \figref{fig:Vsg} and \figref{fig:Vse}, a normalized coupling strength $\Omega_{\rm R}/\omega_c = 0.15$ is sufficient to break one-, two- and three-photon loops, converting virtual photons into real ones with probability close to one. This value of the coupling strength is roughly equal to the experimentally demonstrated values in  circuit-QED systems \cite{Niemczyk2010}. Very recently, by making use of the macroscopic magnetic dipole moment of a flux qubit, large zero-point-fluctuation current of an LC oscillator, and large Josephson inductance of a coupler junction, quantum circuits  where $\Omega_{\rm R}/\omega_{\rm eg}$ ranges from 0.72 to 1.34, have been realized \cite{Yoshihara2016}.

\section{Discussion}

The results presented here show that the USC regime of cavity QED can be exploited to observe, in a direct way, how interactions dress observed particles by a cloud of virtual particles. Such particle dressing is a general feature of quantum field theory and many-body quantum systems. We have shown that, by applying external electromagnetic pulses of suitable amplitude and frequency, each virtual photon enriching a physical excitation can be converted into a physical observable photon. In this way, the hidden relationship between the bare and physical excitations can be unravelled and becomes experimentally testable. 

We have shown that the Feynman diagrams describing the photon emission from a physical excitation are closely linked to the loop diagrams describing the energy correction of a physical excitation induced by the interaction. These loop or bubble diagrams describe internal processes where virtual photons are created and reabsorbed. The diagrams describing the conversion of virtual photons, dressing a physical excitation, into real ones can be obtained by cutting the loop diagrams describing energy corrections and taking the first half as corresponding to the creation of  photons. Moreover, the stimulated reabsorption of real photons into the physical excitation, converting them to virtual photons, corresponds to the second half of the loop diagrams.

We limited our analysis to the dressed vacuum and to a one-particle state. It can be easily extended to study higher-energy excitations. Moreover, we considered only processes up to second-order perturbation theory. The present analysis can be generalized to describe higher-order processes, involving more than three photons, which can take place if the light-matter interaction is sufficiently strong \cite{Yoshihara2016}. 

The most promising candidates for an experimental realization of the proposed stimulated conversion effects are superconducting quantum circuits and intersubband quantum-well polaritons. In particular, phase-biased flux qubits can reach the USC regime in circuit QED \cite{Bourassa2009}, as has been shown in experiments \cite{Niemczyk2010, Baust2014,Yoshihara2016}. By adjusting the externally applied magnetic flux, these artificial atoms can acquire both the quantized level structure and the transition matrix elements required for the observation of the stimulated emission and reabsorption of virtual particles \cite{Stassi2013}. The USC regime can also be reached for intersubband transitions in undoped quantum wells \cite{Gunter2009}. In this system, an optical resonator in the terahertz spectral range is resonantly coupled to transitions between the two-lowest energy conduction subbands of a large number of identical undoped quantum wells. In this case, the upper valence subband plays the role of the lowest energy state $\ket{s}$ (see \figref{fig:SchematicAndEnergyLevels}). Ultrafast optical pulses can induce transitions between the valence and conduction subbands prompting the conversion from virtual to real photons and \emph{vice versa}. Such experiments, being able to look inside the loops of Feynman diagrams, would provide deep insight into fundamental aspects of interaction processes in QFT.

\section*{Acknowledgements}
This work is partially supported by the RIKEN iTHES Project, the MURI Center for Dynamic Magneto-Optics via the AFOSR award number FA9550-14-1-0040,
the IMPACT program of JST, a Grant-in-Aid for Scientific Research (A), and from the MPNS COST Action MP1403 Nanoscale Quantum Optics. A.F.K. acknowledges support from a JSPS Postdoctoral Fellowship for Overseas Researchers.
\section*{Supplementary Material}
In this Supplementary Material, we first restate some properties of the Rabi model and its diagrammatic representation, expanding on the discussion in the main text. We then proceed to explicitly calculate analytically the second-order correction to the lowest energy eigenvalues and comparing them to full numerical calculations. We also calculate matrix elements associated with the external drive used to stimulate the emission and reabsorption of the virtual particles dressing the excitations in the system.

\subsection{Hamiltonian and basic diagrams}

The interaction Hamiltonian of the Rabi model is 
\begin{equation}
\hat V = \Omega_{\rm R} \left(\hat a^\dagger +\hat a \right) \sx,
\label{VR}
\end{equation}
where $\Omega_{\rm R}$ is the coupling strength and $\sx = \sp + \sm = \ketbra{e}{g} + \ketbra{g}{e}$. Referring to the case $\omega_{\rm c} \approx \omega_{eg} \equiv \omega_e - \omega_g$, the interaction Hamiltonian can be separated into a resonant and a nonresonant contribution: $\hat V = \hat V_{\rm r} + \hat V_{\rm nr}$, where $\hat V_{\rm r} = \Omega_{\rm R} \left(\hat a^\dag \sm + \hat a \sp \right)$, and $\hat V_{\rm nr} = \Omega_{\rm R} \left(\hat a^\dag \sp + \hat a \sm \right)$. This interaction term has a structure which is very similar to that of the QED interaction potential, although it is simpler. The Rabi model can be viewed as a prototypical QED system where there is only one photon mode and a two-state electron. Therefore, we expect that the Feynman diagrams for the Rabi Hamiltonian will be a simplified version of the  QED diagrams. 

\begin{figure}
	\includegraphics[width=120 mm]{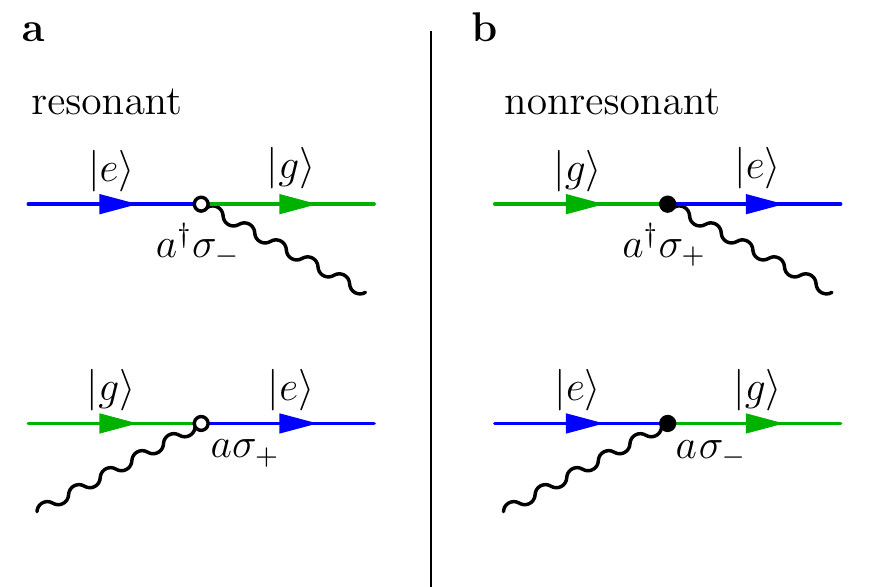}
	\caption{Diagrams corresponding to the four terms in the interaction Hamiltonian of the Rabi model. The horizontal lines represent the qubit states and the wavy lines the cavity photons. (a) Diagrams corresponding to those terms in the interaction Hamiltonian that conserve energy when $\omega_{\rm c} = \omega_{eg}$. (b) Diagrams for the terms $\propto \hat a^\dag \sp$ and $\hat a \sm$, which do not conserve neither energy nor the number of excitations. Their elimination corresponds to the RWA. \label{fig:FourVerticesRabi}}
\end{figure}

As in QED, there is only one vertex type with three lines: One wavy (photonic) line, one solid line with an incoming arrow, and one solid line with an outgoing arrow. The vertices (of the same type) corresponding to the four terms in the interaction Hamiltonian are displayed in \figref{fig:FourVerticesRabi}.
The upper diagram in \figref{fig:FourVerticesRabi}a describes the spontaneous emission process and the lower one the absorption process. Starting from these four building blocks, it is possible to describe higher-order processes as in QED. However, in cavity QED there are processes that are not described in a complete way by Feynman diagrams directly derived from this form of the interaction Hamiltonian. Specifically, the presence of a resonator supporting discrete modes opens up the possibility of observing processes involving more than one photon in the same mode. Stimulated emission, the process underlying laser action, is one of these. It is a one-photon process $\ket{e,n} \to \ket{g,n+1}$, where, however, the $n$ photons in the initial state stimulate the downward transition of the atom, affecting the transition rate which becomes proportional to $n+1$. The Feynman diagram describing the process is the same one describing spontaneous emission ($n=0$), shown in \figref{fig:FourVerticesRabi}a. However, the transition rate for stimulated emission is $n+1$ times larger than that of spontaneous emission. Hence the Feynman diagram in the absence of additional rules is not able to determine uniquely the transition amplitude for this process. 

\begin{figure}
	\includegraphics[width=120 mm]{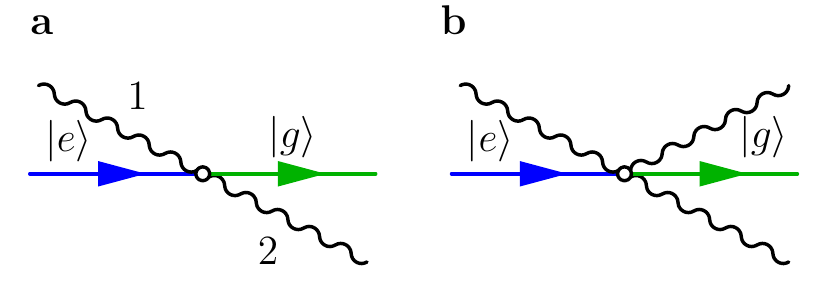}
	\caption{Diagrams corresponding to the the transition $\ket{e,1} \rightarrow \ket{g,2}$. (a) Diagram with one incoming and one outcoming wavy line, each labelled respectively with the number of photons involved in the process. (b) Diagram for the same process, but in this case each wavy line represents one single photon. \label{fig:DiagramsE1G2}}
\end{figure}

A possible solution is to expand the photon creation and destruction operators in \eqref{VR} in the Fock basis. The resulting interaction operator is
\begin{equation}
\hat V = \Omega_{\rm R} \sum_{n=0}^\infty \left(\hat \alpha_+^{(n)} + \hat \alpha_-^{(n)} \right) \left( \sp + \sm \right),
\label{VR2}
\end{equation}
where $\hat \alpha_+^{(n)} = \hat a^\dag \ketbra{n}{n} = \sqrt{n+1}\ketbra{n+1}{n}$, and $\hat \alpha_-^{(n)} = \hat a \ketbra{n}{n} = \sqrt{n}\ketbra{n-1}{n}$ (notice that $\hat \alpha_-^{(0)}=0$). This form of the interaction Hamiltonian consists of a sum of products of (upward or downward) atomic and photonic transition operators; thus photonic and atomic transitions are treated on an equal footing. In this case, each vertex is associated to two transition operators. For example, the vertex describing the transition $\ket{e,1} \rightarrow \ket{g,2}$ is shown in \figref{fig:DiagramsE1G2}a: the wavy lines describe the incoming and the outgoing photon states, while the continuous line with the arrows describes the incoming and outgoing electronic states. The vertex in \figref{fig:DiagramsE1G2}a describes a stimulated emission process. Since the photon wavy lines are labelled by the photon number, an alternative (perhaps more visual) way to draw diagrams is to drop the photon label and draw a wavy line for each incoming or outgoing photon line as shown in \figref{fig:DiagramsE1G2}b. In this case, the vertices will have $n_{\rm in}=n$ incoming and $n_{\rm out}=n \pm 1$ outgoing wavy lines. Each vertex (full/empty circle) contributes with a factor $\sqrt{n}\abs{\hat V_{\rm r/nr}}$, where $n = \max (n_{\rm in},n_{\rm out})$.

\begin{figure}
	\includegraphics[width=80 mm]{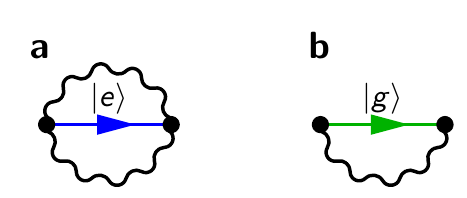}
	\caption{Examples of diagrams corresponding to the Green's Function. (a) Diagram for $\brakket{e,2}{\hat G_0}{e,2}$. (b) Diagram for $\brakket{g,1}{\hat G_0}{g,1}$. \label{fig:GreensFunctionDiagrams}}
\end{figure}

The Green's function for the system in the absence of interaction, 
\be
\hat G^{(n)}_q(z) \equiv \brakket{q,n}{\hat G_0}{q,n} = \frac{1}{z - (\omega_{qg} + n\omega_{\rm c})},
\ee
where $\omega_{qg}=\omega_{q}-\omega_{g}$, with $q=e,g$, corresponds to a loop diagram with $n$ wavy lines and one straight arrow. In \figref{fig:GreensFunctionDiagrams}, we show the two loop diagrams corresponding to $G^{(2)}_e$ and $G^{(1)}_g$.

\subsection{Second-order correction to the energy eigenvalues}	

The well-known second-order correction to the $n$th energy eigenvalue is
\be
\Delta E_n^{(2)}= \sum\nolimits^{\prime}_{k} \frac{\abssq{\brakket{E_k}{\hat V}{E_n}}}{E^0_n-E^0_k},
\label{a1}
\ee
where the prime in the summation means that the values $k=n$ have to be excluded. For the first-order correction to the eigenfunction we have
\begin{equation}\label{K1}
\ket{E_n^{(1)}}= \sum\nolimits^{\prime}_{k} \frac{\brakket{E_k}{\hat V}{E_n}}{E^0_n-E^0_k} \ket{E_k}.
\end{equation}
Following \cite{Konishi2009}, defining the projection operator onto the space orthogonal to $\ket{n}$, $\hat Q_n= \hat 1 - \ketbra{n}{n}$, \eqref{K1} becomes
\begin{equation}
\ket{E_n^{(1)}} = \hat Q_n \hat G(E^{0}_n) \hat Q_n \hat V \ket{E_n},
\label{a10}
\end{equation}
where 
\begin{equation}
\hat G_0(E^{0}_n) = \frac{1}{E^0_n - \hat H_0}
\label{g1}
\end{equation}
is the unperturbed Green's function calculated for $E^0_n$, the unperturbed eigenenergy of the system.

Using the definition of the Green's function from \eqref{g1} and the projection operators, \eqref{a1} becomes
\begin{equation}
\Delta E_n^{(2)} = \brakket{E_k}{\hat V \hat Q_n \hat G(E_n) \hat Q_n \hat V}{E_n}.
\label{a2}
\end{equation}
We apply these results to the JC Hamiltonian perturbed by the nonresonant potential $\hat V_{\rm nr}$. In this case, the unperturbed Hamiltonian $\hat H_0$ becomes $\hat H_{JC} = \hat H_0 + \hat V_{\rm r}$, whose eigenvalues and eigenstates are ${\cal E}^{\pm}_n$ and $\ket{{\cal E}^{\pm}_n}$, respectively. We have
\begin{equation}
\ket{{\cal E}^+_n} = {\cal C}_{n} \ket{g,n} + {\cal S}_{n} \ket{e,n-1} \qquad \ket{{\cal E}^-_n} = - {\cal S}_{n} \ket{g,n} + {\cal C}_{n} \ket{e,n-1}.
\end{equation}

The action of the nonresonant potential on these eigenstates is
\be
\hat V_{\rm nr} \ket{{\cal E}^+_n} = \Omega_{\rm R}({\cal C}_n \ket{e,n+1} + {\cal S}_n \ket{g,n-2})
\ee
and
\begin{equation}
\hat V_{\rm nr} \ket{{\cal E}^-_n} =\Omega_{\rm R}( - {\cal S}_n \ket{e,n+1} + {\cal C}_n \ket{g,n-2}).
\end{equation}
From the last two equations, we deduce that the nonresonant potential $\hat V_{\rm nr}$ determines transitions from the subspace $n$ (spanned by ${\cal E}^{\pm}_n$) to $(n+2)$ or $(n-2)$ subspaces. As a consequence, we have
\be
\hat Q_n \hat V_{\rm nr} \ket{{\cal E}^\pm_n} = \hat V_{\rm nr} \ket{{\cal E}^\pm_n}.
\ee
Owing to this property, \eqref{a2} becomes
\begin{equation}
\Delta^{^{\pm}(2)} = \Delta {\cal E}_n^{\pm(2)} = \brakket{{\cal E}^{\pm}_n}{\hat V_{\rm nr} \hat G({\cal E}^{\pm}_n) \hat V_{\rm nr}}{{\cal E}^{\pm}_n},
\label{a3}
\end{equation}
where
\be
\hat G({\cal E}^{\pm}_n) = \left({\cal E}^{\pm}_n - \hat H_{JC} \right)^{-1} = \left( {\cal E}^{\pm}_n - {\hat H_0} - \hat V_{\rm r} \right)^{-1}
\ee
is the JC Green's function. Equation (\ref{a3}) can be easily calculated exploiting the matrix elements of the JC Green's function by using the JC eigenstates. We do not follow this procedure because our scope is to show, through a diagrammatic analysis, the structure of the virtual processes that contribute to such a correction. For this purpose, we exploit the Dyson equation for the JC Green's function, considering now the resonant potential $\hat V_{\rm r}$ as the perturbation, and the Green's function in the absence of interaction $\hat G_0({\cal E}_n) = \left({\cal E}_n - H_0 \right)^{-1}$: $\hat G = \hat G_0 + \hat G_0 \hat V_{\rm r} \hat G_0 + \dots$. Equation (\ref{a3}) can thus be expanded as
\begin{equation}
\Delta^{^{\pm}(2)}_n = \brakket{{\cal E}^{\pm}_n}{\hat V_{\rm nr} \hat G_0({\cal E}^{\pm}_n) \hat V_{\rm nr}}{{\cal E}^{\pm}_n} + \brakket{{\cal E}^{\pm}_n}{\hat V_{\rm nr} \hat G_0({\cal E}^{\pm}_n) \hat V_{\rm r} \hat G_0({\cal E}^{\pm}_n) \hat V_{\rm nr}}{{\cal E}^{\pm}_n} + \dots
\label{delta0exp}
\end{equation}
%


The lowest-order (second-order) correction to the ground state $\ket{g,0}$ energy due to the nonresonant potential $\hat V_{\rm nr}$, using \eqref{delta0exp}, can be expressed as
\begin{equation}
\Delta_0^{(2)} =\brakket{g,0}{\hat V_{\rm nr} \hat G(E_0) \hat V_{\rm nr}}{g,0} = \brakket{g,0}{\hat V_{\rm nr} \hat G_0 \hat V_{\rm nr}}{g,0} + \brakket{g,0}{\hat V_{\rm nr} \hat G_0 \hat V_{\rm r} \hat G_0 \hat V_{\rm nr}}{g,0} + \dots
\label{E}
\end{equation}
By using the identity operator and exploiting the explicit expression of $\hat V_{\rm nr}$, \eqref{E} can be expressed as
\begin{equation}	
\Delta_0^{(2)} = \brakket{g,0}{\hat V_{\rm nr}}{e,1} \brakket{e,1}{\hat G}{e,1} \brakket{e,1}{\hat V_{\rm nr}}{g,0} = \Omega_{\rm R}^2 \brakket{e,1}{\hat G}{e,1}.
\label{Dy1}
\end{equation}

In order to calculate $\Delta_0^{(2)}$, we observe that 
\be
\brakket{g,0}{\hat V_{\rm nr}}{e,1} = \brakket{e,1}{\hat V_{\rm nr}}{g,0} = \Omega_{\rm R}.
\ee
The remaining term, $\brakket{e,1}{\hat G}{e,1}$, is a convergent geometric series that is calculated in  \secref{sec:CalculationGMatrixElements}. We obtain
\begin{equation}	
\Delta_0^{(2)} = \frac{\Omega_{\rm R}^2}{2\omega_c}\frac{1}{\Omega_{\rm R}^2/2\omega_c^2 - 1}.
\label{Dy10}
\end{equation}
For $\Omega_{\rm R}/\omega_{\rm c}<1$, $\Delta^{(2)}_1$ can be approximated to second order in $\Omega_{\rm R}$:
\begin{equation}
\Delta^{(2)}_0 \approx - \frac{\Omega_{\rm R}^2}{2\omega_c}
\end{equation}

\begin{figure}
	\centering
	\includegraphics[width=\textwidth]{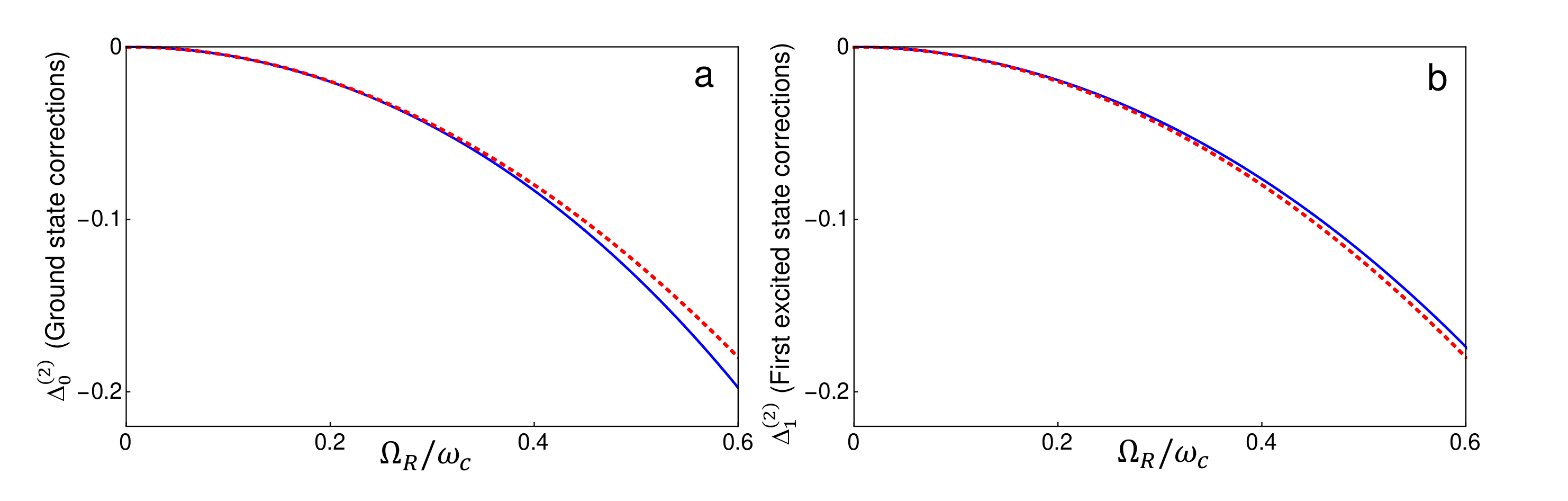} 
	\caption{Comparison between the exact (numerical)  and approximated (diagrammatic) calculation of the correction terms (a) for the ground-state energy and (b) for the first excited state. The blue continuous lines describe the numerical calculation, while the red dotted lines describe the approximate one. \label{fig:ComparingEnergyCorrections}} 
\end{figure}

In \figref{fig:ComparingEnergyCorrections}, we show the comparison between the exact (numerical) and approximated (diagrammatic) calculation of the correction term to the ground state energy.

This approach can also be applied to the excited states. We consider the first excited state. Using \eqref{a3}, we are able to calculate the correction up to second order in the potential $\hat V_{\rm sg}$ to $E_1$; we have
\begin{equation}
\Delta^{(2)}_1 = \brakket{{\cal E}^-_1}{\hat V_{\rm nr} \hat G({\cal E}^-_1) \hat V_{\rm nr}}{{\cal E}^-_1} = {\cal S}^2_1 \brakket{g,1}{\hat V_{\rm nr} \hat G({\cal E}^-_1) \hat V_{\rm nr}}{g,1}.
\label{delta01}
\end{equation}
Observing that $\hat V_{\rm nr} \ket{e,0} = 0$, $\hat V_{\rm nr} \ket{g,1} = \sqrt{2}\Omega_{\rm R} \ket{e,2}$, and $\ket{{\cal E}^-_1} = - {\cal S}_1 \ket{g,1} +{\cal C}_1  \ket{e,0}$, we have
\begin{equation}
\hat V_{\rm nr} \ket{{\cal E}^-_1} =- \sqrt{2}{\cal S}_1\Omega_{\rm R} \ket{e,2}.
\label{delta02}
\end{equation}
Equation (\ref{delta01}) becomes
\bea
\Delta^{(2)}_1 &=& \brakket{{\cal E}^-_1}{\hat V_{\rm nr} \hat G({\cal E}^-_1) \hat V_{\rm nr}}{{\cal E}^-_1} = {\cal S}^2_1 \brakket{g,1}{\hat V_{\rm nr} \hat G({\cal E}^-_1) \hat V_{\rm nr}}{g,1} \nn\\
&=& 2{\cal S}^2_1\,  \Omega_{\rm R}^2 \brakket{e,0}{\hat G({\cal E}^-_1)}{e,0}.
\label{delta1}
\eea

The energy correction  $\Delta^{(2)}_1$ can be evaluated easily by directly using $\hat G(z)$ or by summing up the infinite contributions arising from the Dyson series, described by the diagrams (see \secref{sec:CalculationGMatrixElements}, \eqref{Se21}). 

In the absence of detuning, we have for the first excited state
\be
\ket{{\cal E}^-_1} = \frac{1}{\sqrt{2}} \left( -\ket{g,1} + \ket{e,0} \right),
\ee
with energy
\be
{\cal E}^-_1 = \omega_{\rm c} - \Omega_{\rm R}.
\ee
We obtain
\be
\Delta^{(2)}_1 =  - \Omega_{\rm R}^2 \frac{	\Omega_{\rm R} + 2\omega_{\rm c}}{4\omega_{\rm c}^2 - 4\Omega_{\rm R}\omega_{\rm c} - 2\Omega_{\rm R}^2}.
\label{delta3ter}
\ee
For $\Omega_{\rm R}/\omega_{\rm c}<1$, $\Delta^{(2)}_1$ can be approximated to second order in $\Omega_{\rm R}$:
\begin{equation}
\Delta^{(2)}_1 \approx - \frac{\Omega_{\rm R}^2}{2\omega_{\rm c}}.
\end{equation}
A comparison between the approximate analytical energy corrections and the corresponding nonperturbative numerical calculations can be found in \figref{fig:ComparingEnergyCorrections}.


\subsection{Additional perturbation $\hat V_{\rm sg}$ allowing transitions from $\ket{g}$ to $\ket{s}$}

Using \eqref{a1}, the correction to the JC eigenstate $\ket{{\cal E}_n}$ up to the first order in the nonresonant potential  becomes,
\begin{equation}
\ket{E_n^{(1)}} = \hat Q_n \hat G({\cal E}_n) \hat V_{\rm nr} \ket{{\cal E}_n}.
\label{a4}
\end{equation}

We now consider the direct excitation of the artificial atom by applied electromagnetic pulses, described by the Hamiltonian
\begin{equation}
\hat H_{\rm d} = {\cal E}(t)(\hat V_{sg} +\hat V_{se}),
\label{Hdsupp}
\end{equation}
where $\hat V_{sg} = \mu_{sg} (\ketbra{g}{s} + \ketbra{s}{g})$, $\hat V_{se} = \mu_{\rm se} (\ketbra{e}{s} + \ketbra{s}{e})$, and $\mu_{\rm sg}$ and $\mu_{\rm se}$ are the dipole moments (here assumed to be real) for the transitions $\ket{s} \leftrightarrow \ket{g}$ and $\ket{s} \leftrightarrow \ket{e}$, respectively.

First, we consider the case with the system prepared in the state $\ket{{\cal E}_0} = \ket{g,0}$. The time-dependent perturbation can induce additional transitions whose rate can be evaluated with the Fermi golden rule. 

In the absence of the counter-rotating interaction terms $\hat V_{\rm nr}$, $\hat H_{\rm d}$ can induce only zero-cavity-photon transitions $\ket{g,0} \leftrightarrow \ket{s,0}$. When including the counter-rotating terms,
additional transitions are activated. For example, the transition $\ket{E_0} \leftrightarrow \ket{s,2}$ acquires a nonzero matrix element $\brakket{s,2}{\hat V_{\rm sg}}{E_0}$, where $\ket{E_0}$ is the lowest energy state of the Rabi Hamiltonian.
It can be calculated perturbatively in $\hat V_{\rm nr}$, approximating $\ket{E_0}$ to  first order in $\hat V_{\rm nr}$ (see \eqref{a10}):
\be	
\ket{E_0} \simeq \ket{g,0} +\hat G({\cal E}_0) \hat V_{\rm nr} \ket{g,0}.
\ee
We obtain
\be
\brakket{s,2}{\hat V_{\rm sg}}{E_0} = \brakket{s, 2}{\hat V_{\rm sg} \hat G({\cal E}_0) \hat V_{\rm nr}}{g,0} = \Omega_{\rm R}\mu_{\rm sg} \brakket{g,2}{\hat G({\cal E}_0)}{e,1}.
\ee

\begin{figure}
	\centering
	\includegraphics[width=\textwidth]{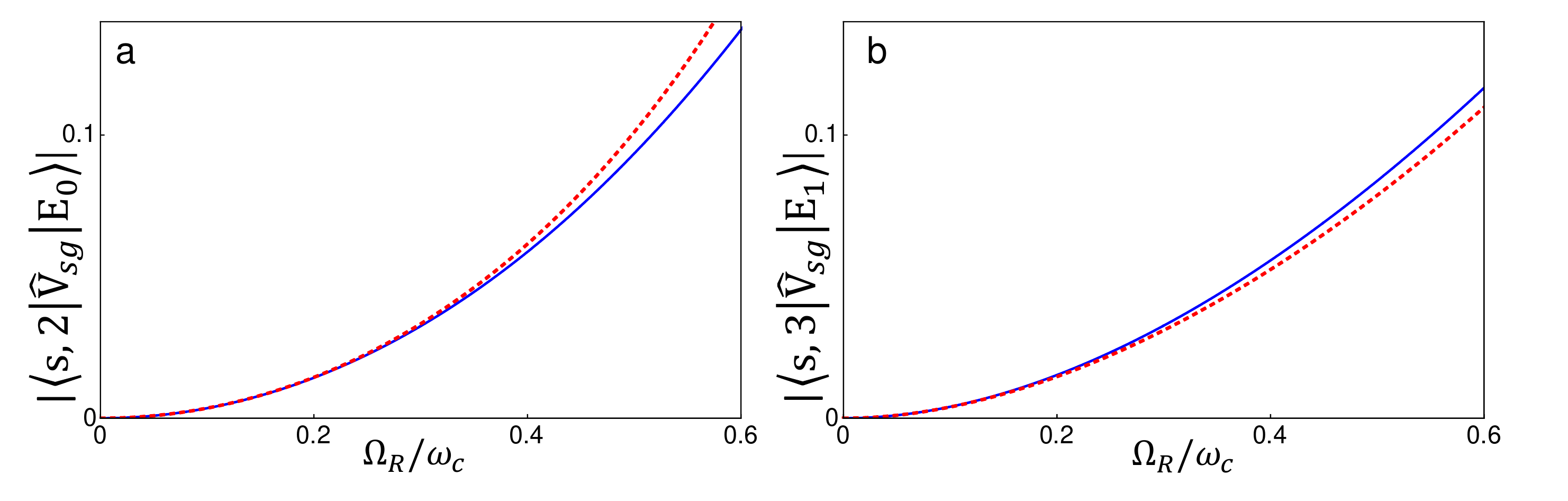} 
	\caption{(a) Comparison between the exact (numerical, blue continuous curve) and approximated (diagrammatic, red dotted curve) calculation of the transition element $\brakket{s,2}{\hat V_{\rm sg}}{E_0}$ between the state $\ket{s,2}$ (where $\ket{s}$ is now the real ground state) and $\ket{E_0}$ (the cavity-dressed ground state) due to the additional $\hat V_{\rm sg}$ potential as a function of the coupling parameter normalized to the cavity resonance frequency.
		(b) Comparison between the exact (numerical) and approximated (diagrammatic) calculation of the transition element $\brakket{s,3}{\hat V_{\rm sg}}{E_1}$ between the state $\ket{s,3}$ (where $\ket{s}$ is now the real ground state) and $\ket{E_1}$ (the cavity-dressed first excited state) due to the additional $\hat V_{\rm sg}$ potential as a function of the coupling parameter, normalized to the cavity resonance frequency. This comparison confirms the validity of the approximate diagrammatic approach. \label{fig:ComparingTransitionMatrixElements}} 
\end{figure}

The corresponding Dyson series is
\bea
\brakket{g,2}{\hat G({\cal E}_0)}{e,1} &=& \brakket{g,2}{\hat G_0({\cal E}_0)}{e,1} + \brakket{g,2}{\hat G_0({\cal E}_0) \hat V_r \hat G_0({\cal E}_0)}{e,1} + \ldots \nn\\
&=& \frac{\sqrt{2} \Omega_{\rm R}}{({\cal E}_0 - 2\omega_c)^2 - 2\Omega_{\rm R}^2} = \frac{\sqrt{2} \Omega_{\rm R}}{4\omega_c^2 - 2\Omega_{\rm R}^2}.
\eea
In \figref{fig:ComparingTransitionMatrixElements} we compare the exact (numerical) and approximated (diagrammatic) calculation of this matrix element.

We now consider the case with the system  prepared in the state $\ket{{\cal E}^-_1} = \dfrac{1}{\sqrt{2}} (-\ket{g,1} + \ket{e,0})$, whose energy is ${\cal E}^-_1= \omega_c - \Omega_{\rm R}$. The time-dependent perturbation can induce additional transitions whose rate can be evaluated with the Fermi golden rule. In the presence of $\hat V_{\rm nr}$, additional transitions, such as $\ket{E_1} \leftrightarrow \ket{s,3}$, become activated. The matrix element for this transition is $\brakket{s,3}{\hat V_{\rm sg}}{E_1}$. It can be calculated perturbatively in $\hat V_{\rm nr}$, approximating $\ket{E_1}$ to first order in $\hat V_{\rm nr}$ (see \eqref{a10}):
\be	
\ket{E_1} \simeq \ket{{\cal E}^-_1}+ \hat G ({\cal E}^-_1) \hat V_{\rm nr} \ket{{\cal E}^-_1}.
\ee
Observing that 
\be
\hat V_{\rm nr} \ket{{\cal E}^-_1} =- \frac{1}{\sqrt{2}} \left(\hat V_{\rm nr} \ket{g,1} - \hat V_{\rm nr} \ket{e,0} \right) =- \Omega_{\rm R} \ket{e,2},
\ee
where we have used the relations $\hat V_{\rm nr} \ket{g,1} = \sqrt{2}\Omega_{\rm R} \ket{e,2}$ and $\hat V_{\rm nr} \ket{e,0} = 0$,
we obtain
\be
\brakket{s,3}{\hat V_{\rm sg}}{E_1} = \brakket{s,3}{\hat V_{\rm sg} \hat G({\cal E}^-_1) \hat V_{\rm nr}}{{\cal E}^-_1} = -\Omega_{\rm R}\mu_{\rm sg} \brakket{g,3}{\hat G ({\cal E}^-_1)}{e,2}.
\ee
The corresponding Dyson series is (see \secref{sec:CalculationGMatrixElements} and \eqref{Se22})
\bea
\brakket{g,3}{\hat G ({\cal E}^-_1)}{e,2} &=& \brakket{g,3}{\hat G_0({\cal E}^-_1)}{e,2} + \brakket{g,3}{\hat G_0({\cal E}^-_1) \hat V_r \hat G_0({\cal E}^-_1)}{e,2} + \ldots \nn\\
&=& \frac{\sqrt{3} \Omega_{\rm R}}{({\cal E}_1 - 3\omega_c)^2 - 3 \Omega_{\rm R}^2} = \frac{\sqrt{3} \Omega_{\rm R}}{( \Omega_{\rm R} + 2\omega_c)^2 - 3 \Omega_{\rm R}^2}.
\eea



\subsection{Additional perturbation $\hat V_{\rm se}$ allowing transitions from $\ket{e}$ to $\ket{s}$}

We now consider a situation analogous to that analyzed before. In this case, a transition $\ket{s} \leftrightarrow \ket{e}$ is detuned at a much higher energy than the cavity resonance. The part of the time-dependent potential inducing the $\ket{s} \leftrightarrow \ket{e}$ transitions is $\hat V'(t) = {\cal E}(t)\hat V_{se}$. In the absence of the counter-rotating interaction terms $\hat V_{\rm nr}$, $\hat V'(t)$ can induce zero-cavity-photon transitions $\ket{e,0} \leftrightarrow \ket{s,0}$.


In the presence of $\hat V_{\rm nr}$, additional transitions, such as $\ket{E_0} \leftrightarrow \ket{s,1}$, can be activated. The matrix element for this transition is $\brakket{s,1}{\hat V_{\rm se}}{E_0}$. It can be calculated perturbatively in $\hat V_{\rm nr}$, approximating $\ket{E_0}$ to the first order in $\hat V_{\rm nr}$ (see \eqref{a10}):
\be	
\ket{E_0} \simeq \ket{g,0}+ \hat G ({\cal E}_0) \hat V_{\rm nr} \ket{g,0}.
\ee
We have
\be
\brakket{s,1}{\hat V_{\rm se}}{E_0} = \brakket{s,1}{\hat V_{\rm se} \hat G_0({\cal E}_0) \hat V_{\rm nr}}{g,0} = \Omega_{\rm R}\mu_{\rm se} \brakket{e,1}{\hat G_0({\cal E}_0)}{e,1}.
\label{e0cut}
\ee
Exploiting the Dyson series for the Green's function, we obtain
\be	
\brakket{e,1}{\hat G ({\cal E}_0)}{e,1} = \brakket{e,1}{\hat G_0({\cal E}_0)}{e,1} + \brakket{e,1}{\hat G_0({\cal E}_0) \hat V_r \hat G_0({\cal E}_0)}{e,1} + \dots
\ee
In this series, owing to the nature of the resonant potential, only the odd terms are non-negligible; we have 
\be
\brakket{e,1}{\hat G ({\cal E}_0)}{e,1} = \brakket{e,1}{\hat G_0({\cal E}_0)}{e,1} + \sum_{n=1}^{\infty} \brakket{e,1}{\hat G_0({\cal E}_0) \left(\hat V_r \hat G_0({\cal E}_0) \right)^{2n}}{e,1}.
\ee
The diagrammatic analysis of this process is shown in Fig.~5 of the main part  of the paper. Using the results of \secref{sec:CalculationGMatrixElements}, the Dyson series calculation of \eqref{e0cut} gives on resonance
\be
\brakket{s,1}{\hat V_{\rm se}}{E_0} = \Omega_{\rm R}\mu_{\rm se} \brakket{e,1}{\hat G({\cal E}_0)}{e,1} = \mu_{\rm se}\frac{\Omega_{\rm R} \omega_c }{\Omega_{\rm R}^2 - 2\omega_c^2}
\label{Se2es}
\ee
%


\subsection{Calculation of the $\hat G$ matrix elements using the Dyson equation}
\label{sec:CalculationGMatrixElements}

Here we perform all the calculations for the determination of the correction to the self-energy to second order in $\hat V_{\rm nr}$. We have to sum all the elements of the infinite series. The generic matrix element is
\be
\brakket{e,1}{\hat G_0(z) [\hat V_r\hat G_0(z)]^n}{e,1}.
\ee
We observe that when $\hat V_r$ appears an odd number of times, then the matrix element will be zero, {i.e.},
\be
\brakket{e,1}{\hat G_0(z) [\hat V_r\hat G_0(z)]^{2n+1}}{e,1} = 0 \qquad \mbox{for}\: n>0.
\label{Se1}
\ee
Hence we may perform the calculation for the self-energy considering only the even-power terms. In addition, we observe that
\be
\brakket{g,n+1}{\hat V_r}{e,n} = \brakket{e,n}{\hat V_r}{g,n+1} = \sqrt{n+1}\, \Omega_{\rm R}.
\ee
We then obtain
\bea
\brakket{e,1}{\hat G}{e,1} &=& \sum_{n=0}^{\infty} \brakket{e,1}{\hat G_0(z) \left[\hat V_r\hat G_0(z)\right]^{2n}}{e,1} \nn\\
&=& \brakket{e,1}{\hat G_0(z)}{e,1} \sum_{n=0}^{\infty} \brakket{e,1}{\left[\hat V_r\hat G_0(z)\right]^{2n}}{e,1} \nn\\
&=& \brakket{e,1}{\hat G_0(z)}{e,1} \sum_{n=0}^{\infty}\left| \brakket{e,1}{\hat V_r}{g,2}\right|^{2n} \left( \brakket{e,1}{\hat G_0(z)}{e,1} \right)^n \left( \brakket{g,2}{\hat G_0(z)}{g,2} \right)^n \nn\\
&=& \brakket{e,1}{\hat G_0(z)}{e,1} \sum_{n=0}^{\infty} (\sigma)^n,
\eea
where
%
%
\bea
\sigma &=& \abssq{\brakket{e,1}{\hat V_r}{g,2}}
\brakket{e,1}{\hat G_0(z)}{e,1}
\nn\\
&=&\Omega_{\rm R}^2 \brakket{e,1}{\hat G_0(z)}{e,1}
\brakket{g,2}{\hat G_0(z)}{g,2}
\eea
Following \cite{Economou1984}, the Green's function operator relative to a generic differential operator $\hat{L}$  satisfies the relation
\be
\left[ z - \hat{L} \right] \hat{G}(z) = \hat{I},
\ee
with $z$ a convenient parameter. Therefore,  for the cases $\hat L=\hat H$ and $\hat L=\hat H_0$, we have, respectively:
\be
\hat G(z) = \left[ z - \hat H \right]^{-1}, \quad \hat G_0(z) = \left [ z - \hat H_0 \right]^{-1}.
\ee
%

In our calculation, we choose $z={\cal E}_0$ (hence $\hat G_0(z)$ is the resolvent of the free Hamiltonian eigenvalue problem):
\bea
\brakket{e,1}{\hat G(z)}{e,1} &=& \brakket{e,1}{\hat G_0(z)}{e,1} \frac{1}{1-\sigma} = \frac{ \left (\brakket{g,2}{\hat G_0(z)}{g,2} \right)^{-1}}{ \left( \brakket{e,1}{\hat G_0(z)}{e,1} \right)^{-1} \left (\brakket{g,2}{\hat G_0(z)}{g,2} \right)^{-1} - 2\Omega_{\rm R}^2} \nn\\
&=& 	\frac{	(z - 2\omega_c)}{(z - \omega_{eg} - \omega_c) (z - 2\omega_c) - 2\Omega_{\rm R}^2}
\label{Se2}
\eea	

For completeness, we now calculate  the other non-zero matrix elements:
\bea
\brakket{e,1}{\hat G(z)}{g,2} &=& \sum_{n=0}^{\infty} \brakket{e,1}{\hat G_0(z) \left[\hat V_r\hat G_0(z) \right]^{2n+1}}{g,2} \nn\\
&=& \brakket{g,2}{\hat G_0(z)}{g,2} \brakket{e,1}{\hat G_0(z)}{e,1} \brakket{e,1}{\hat V_r}{g,2} \sum_{n=0}^{\infty} \brakket{e,1}{\left[ \hat V_r\hat G_0(z) \right]^{2n}}{e,1} \nn\\
&=& \brakket{g,2}{\hat G_0(z)}{g,2} \brakket{e,1}{\hat V_r}{g,2} \brakket{e,1}{\hat G(z)}{e,1} \nn\\
&=& \brakket{g,2}{\hat G_0(z)}{g,2} \brakket{e,1}{\hat V_r}{g,2} \brakket{e,1}{\hat G_0(z)}{e,1} \sum_{n=0}^{\infty}(\sigma)^n \nn\\
&=& \brakket{e,1}{\hat G_0(z)}{e,1} \brakket{g,2}{\hat G_0(z)}{g,2} \frac{\sqrt{2}\Omega_{\rm R}}{1-\sigma} \nn\\
&=& \frac{\sqrt{2}\Omega_{\rm R}}{ \left( \brakket{e,1}{\hat G_0(z)}{e,1} \right)^{-1} \left( \brakket{g,2}{\hat G_0(z)}{g,2} \right)^{-1} - 2\Omega_{\rm R}^2} \nn\\
&=& \frac{\sqrt{2}\Omega_{\rm R}}{(z - \omega_{eg} - \omega_c) (z - 2\omega_c) - 2\Omega_{\rm R}^2}
\eea
and
\bea
\brakket{g,2}{\hat G(z)}{g,2} &=& \brakket{g,2}{\hat G_0(z)}{g,2} \frac{1}{1-\sigma} \nn\\
&=& \frac{ \left( \brakket{e,1}{\hat G(z)}{e,1} \right)^{-1}}{ \left( \brakket{e,1}{\hat G_0(z)}{e,1} \right)^{-1} \left( \brakket{g,2}{\hat G_0(z)}{g,2} \right)^{-1} - 2\Omega_{\rm R}^2} \nn\\
&=& \frac{	(z - \omega_{eg} - \omega_c)}{(z - \omega_{eg} - \omega_c) (z - 2\omega_c) - 2\Omega_{\rm R}^2}.
\eea
The generalization of the above matrix elements to all $n$ subspaces is straightforward. Indeed, we can now calculate all the contributions of the self-energy: the Jaynes-Cummings Hamiltonian divides the entire Hilbert space into disjoint 2D subspaces (labelled by $n$) spanned by ($\ket{e,n}$, $\ket{g,n+1}$). For the generic $n$th subspace we have
\bea
\brakket{e,n}{\hat G(z)}{e,n} &=& \frac{(z - (n+1)\omega_c)}{(z - \omega_{eg} - n\omega_c) (z - (n+1)\omega_c) - (n+1)\Omega_{\rm R}^2},
\label{Se21} \\
\brakket{e,n}{\hat G(z)}{g,n+1} &=& \frac{\sqrt{n+1}\Omega_{\rm R}}{(z - \omega_{eg} - n\omega_c) (z - (n+1)\omega_c) - (n+1)\Omega_{\rm R}^2},
\label{Se22} \\
\brakket{g,n+1}{\hat G(z)}{g,n+1} &=& \frac{(z - \omega_{eg} - n\omega_c)}{(z - \omega_{eg} - n\omega_c)(z - (n+1) \omega_c) - (n+1)\Omega_{\rm R}^2}.
\label{Se23}
\eea

\bibliography{FeynmanRefs}

\end{document}